\documentclass[12pt,reqno]{amsart}
\usepackage{amsmath}
\usepackage{latexsym}
\usepackage{amsfonts}
\usepackage{amssymb}
\usepackage{color}
\usepackage{bbm,dsfont}
\usepackage{graphicx}

\newtheorem{proposition}{Proposition}

\theoremstyle{definition}

\newtheorem{example}{Example}
\newtheorem{definition}{Definition}

\newcommand{\real}{\mathbb R} 

\newcommand{\hi}{\mathcal{H}} 
\newcommand{\lh}{\mathcal{L(H)}} 
\newcommand{\trh}{\mathcal{T(H)}} 
\newcommand{\sh}{\mathcal{S(H)}} 
\newcommand{\eh}{\mathcal{E(H)}} 
\newcommand{\ip}[2]{\left\langle\,#1\,|\,#2\,\right\rangle} 
\newcommand{\ket}[1]{|#1\rangle} 
\newcommand{\bra}[1]{\langle#1|} 
\newcommand{\kb}[2]{|#1\rangle\langle#2|} 
\newcommand{\tr}[1]{\textrm{tr}\left[#1\right]} 
\newcommand{\ptr}[1]{\textrm{tr}_1[#1]} 
\newcommand{\ran}{\textrm{ran}} 
\newcommand{\id}{I} 
\newcommand{\nul}{O} 

\newcommand{\meo}{\mathcal{O}} 
\newcommand{\salg}{\mathcal{F}} 

\newcommand{\G}{\mathsf{G}}
\renewcommand{\L}{\mathcal{L}}

\newcommand{\U}{\mathcal{U}} 
\newcommand{\Lu}{\Phi^\mathcal{L}} 

\newcommand{\J}{\mathcal{J}}
\newcommand{\I}{\mathcal{I}}
\newcommand{\E}{\mathcal{E}}

\begin{document}

\title[Coexistence of quantum operations]{Coexistence of quantum operations}

\author[Heinosaari]{Teiko Heinosaari}
\address{Teiko Heinosaari, Niels Bohr Institute, Blegdamsvej 17, 2100 Copenhagen, Denmark}
\email{heinosaari@nbi.dk}

\author[Reitzner]{Daniel Reitzner}
\address{Daniel Reitzner, Institute of Physics, Slovak Academy of Sciences, D\'ubravsk\'a cesta 9, 845 11 Bratislava, Slovakia}
\email{daniel.reitzner@savba.sk}

\author[Stano]{Peter Stano}
\address{Peter Stano, Institute of Physics, Slovak Academy of Sciences, D\'ubravsk\'a cesta 9, 845 11 Bratislava, Slovakia}
\email{peter.stano@savba.sk}

\author[Ziman]{Mario Ziman}
\address{Mario Ziman, Institute of Physics, Slovak Academy of Sciences, D\'ubravsk\'a cesta 9, 845 11 Bratislava, Slovakia and Faculty of Informatics, Masaryk University, Botanick\'a 68a, Brno, Czech Republic}
\email{ziman@savba.sk}

\date{May 29, 2009}

\begin{abstract}
Quantum operations are used to describe
the observed probability distributions and conditional states of the measured
system. In this paper, we address the problem of their joint measurability
(coexistence). We derive two equivalent coexistence criteria. The two most common classes of operations --- L\"uders operations
and conditional state preparators --- are analyzed. It is shown that L\"uders operations
are coexistent only under very restrictive conditions, when the associated
effects are either proportional to each other, or disjoint.
\end{abstract}
\maketitle

\section{Introduction}\label{sec:intro}

One of the basic implications of quantum mechanics is that there exist
incompatible experimental setups. For example, as it was originally
initiated by Werner Heisenberg \cite{Heisenberg27}, the measurements of position and momentum
of a quantum particle cannot be performed simultaneously unless
some imprecisions are introduced \cite{BuHeLa07}. The fact that position and momentum
are not experimentally compatible physical quantities reflects the very properties
of the quantum theory leading to the concept of coexistence.

In general, the coexistence of quantum devices means that they can be
implemented as parts of a single device. Until now, the coexistence relation has been studied among quantum effects and observables; see e.g. \cite{StReHe08} and references therein. However, in addition to observed measurement outcome statistics we can also
end up with a quantum system. Quantum operations and instruments are used to
mathematically describe both: probabilities of the observed measurement outcomes and
conditional states of the measured quantum system post-selected according to
observed outcomes. Compared to an effect, an operation describes a particular result of a quantum measurement on a different level, providing more details about what happened during the measurement. The topic of this paper -- the coexistence of quantum operations -- is thus a natural extension of the previous studies of the coexistence of effects.

Let us now fix the notation and set the problem in mathematical terms.
Let $\hi$ be a complex separable Hilbert space. We denote by $\lh$ and $\trh$ the Banach spaces of bounded operators and trace class operators on $\hi$, respectively. The set of quantum states (i.e. positive trace one operators) is denoted by $\sh$ and the set of quantum effects (i.e. positive operators bounded by the identity) is denoted by $\eh$.

An \emph{operation} $\Phi$ is a completely positive linear mapping
on $\trh$ such that
\begin{equation*}
0 \leq \tr{\Phi(\varrho)} \leq 1
\end{equation*}
for every $\varrho\in\sh$. An operation represents a probabilistic state transformation. Namely, if $\Phi$ is applied on an input state $\varrho$, then the state transformation $\varrho\mapsto\Phi(\varrho)$ occurs with the probability $\tr{\Phi(\varrho)}$, in which case the output state is $\Phi(\varrho)/\tr{\Phi(\varrho)}$. A special class is formed by operations satisfying $\tr{\Phi(\varrho)}=1$ for every state $\varrho\in\sh$; these are called \emph{channels} and they describe deterministic state transformations.

An \emph{instrument} is a device which takes as an input a quantum state, produces a measurement outcome, and conditional to the measurement outcome also produces an output state. Mathematically instrument is represented as an operation valued measure \cite{QTOS76}. To be more precise, let $\meo$ be a set of measurement outcomes and $\salg$ a $\sigma$-algebra of subsets of $\meo$. An instrument $\J$ is a $\sigma$-additive mapping $X\mapsto \J(X)\equiv\J_X$ from $\salg$ to the set of operations on $\trh$.  It is required to satisfy the normalization condition
\begin{equation*}
\tr{\J_\meo(\varrho)}=1 \quad \forall\varrho\in\sh \, ,
\end{equation*}
which means that some state transformation occurs with probability 1.
We denote by $\ran\J$ the range of $\J$, that is,
\begin{equation*}
\ran\J = \{ \J_X \mid X\in\salg \} \, .
\end{equation*}

\begin{definition}\label{def:coexistent}
Two operations $\Phi$ and $\Psi$ are \emph{coexistent} if there exists an instrument $\J$ such that $\Phi,\Psi\in\ran\J$.
\end{definition}

The mathematical coexistence problem we study in this article is to find out whether two given operations are coexistent or not.

This definition is analoguous to definition of effect coexistence.
We recall that two effects are coexistent if there exists an observable (POVM), which
has both these effects in its range \cite{LaPu97}. Since operations are more complex objects than effects, one naturally expects the coexistence relation for operations to be more complicated. On the other hand, the coexistence of effects in general is still an open problem and it has turned out unexpectedly difficult to solve effects coexistence problem even for quite restricted classes of effects. It is therefore not to be expected to find a complete solution of the operation coexistence problem here. Our intention is to formulate the problem properly and clarify the complexity of the tasks involved. As examples, we analyze the coexistence properties of L\"uders operations and conditional state preparators.

The rest of this paper is organized as follows.
In Section \ref{sec:general} we derive two alternative mathematical formulations for the coexistence relation. In Section \ref{sec:rank1} we study a special case of coexistence, called trivial coexistence. We show that pure (i.e. rank-1) operations can be coexistent only under very restrictive conditions. In Section \ref{sec:effects} we compare the coexistence of operations to the coexistence of the associated effects. Finally, Section \ref{sec:discussion} gives our conclusions and an outlook.

\section{General coexistence criteria for operations}\label{sec:general}

We first shortly recall some basic concepts related to quantum operations as these are needed in the formulation of coexistence criteria.
For more details on operations and instruments, we refer the reader to \cite{QTOS76}.

Let $\Phi$ be an operation. The dual mapping $\Phi^\ast:\lh\to\lh$ of $\Phi$ is defined by the
duality formula
\begin{equation}\label{eq:dual}
\tr{\Phi^\ast(R)T} = \tr{R\Phi(T)}\,,
\end{equation}
required to hold for all  $R\in\lh, T\in\trh$.
The dual operation $\Phi^\ast$ describes the same quantum operation as $\Phi$ but in the Heisenberg picture. Setting $R=I$ in Eq.~\eqref{eq:dual} we see that $\Phi$ determines a unique effect $A$ by formula
\begin{equation}\label{eq:operation->effect}
\Phi^\ast(\id)=A \, .
\end{equation}
We often use the subscript notation $\Phi_A$ to emphasize this connection, meaning both $A$ and $\Phi_A$ give rise to the same measurement outcome probabilities. As operations also describe state transformations, it is understandable that the relation $A\to\Phi_A$ is one-to-many rather than one-to-one.

Two useful alternative mathematical descriptions of operations are provided by
\emph{Kraus decomposition} \cite{SEO83} and
\emph{Choi-Jamiolkowski isomorphism} \cite{Choi75,Jamiolkowski72}.
The latter description we formulate only in the case of a finite dimensional Hilbert space.

The Kraus decomposition theorem states that a linear mapping $\Phi$ is an operation if and only if there exists a countable set of bounded operators $\{X_k\}$ such that $\sum_k X_k^* X_k\leq I$ and
\begin{equation}\label{eq:kraus}
\Phi(\varrho) = \sum_k X_k \varrho X_k^\ast
\end{equation}
holds for all $\varrho\in\sh$. Further we also use a short-hand notation of Eq.~(\ref{eq:kraus}) in the form $\Phi= \sum_k X_k \cdot X_k^\ast$. Using a Kraus decomposition \eqref{eq:kraus}
 for $\Phi$ we see that Eq.~\eqref{eq:operation->effect} is equivalent to the condition
\begin{equation*}
A=\sum_{k}X_k^\ast X_k \, .
\end{equation*}

For a fixed operation $\Phi$, the choice of operators $X_k$, referred as Kraus operators, is not unique.
Namely, two sets $\{X_k\}$ and $\{Y_l\}$ determine the same operation if and
only if there are complex numbers $U_{lk}$ such that
$\sum_l \overline{U}_{lj} U_{lk}=\delta_{jk}$ and $Y_l=\sum_k U_{lk}X_k$
\cite{QCQI00}.
When comparing two Kraus decompositions we can always assume that they
have the same number of elements by adding null operators $\nul$
if necessary. With this assumption the numbers $U_{lk}$ form a unitary matrix
$U$.

The essential ingredient of the Choi-Jamiolkowski isomorphism is the so-called maximally entangled state
\begin{equation*}
\psi_+=\frac{1}{\sqrt{d}}\sum_{j} \varphi_j\otimes\varphi_j \, ,
\end{equation*}
where the vectors $\{\varphi_j\}$ form an orthonormal basis of $\hi$ and $d=\dim\hi<\infty$.
We denote by $\I$ the identity operation $\I(\varrho) = \varrho$.
The formulas
\begin{eqnarray}
\Phi&\mapsto&\Xi_\Phi=(\Phi\otimes\I)(\kb{\psi_+}{\psi_+})\\
\Xi&\mapsto&\Phi_\Xi,\quad \Phi_\Xi(\varrho)=d\ptr{(\varrho^T\otimes I)\Xi}
\end{eqnarray}
determine a one-to-one Choi-Jamiolkowski mapping (and its inverse) between
linear mappings $\Phi$ on $\trh$ and operators $\Xi$ on $\hi\otimes\hi$.
The transposition is with respect to the orthonormal basis $\{\varphi_j\}$
used in the definition of the maximally entangled state $\psi_+$.

If $\Phi$ is an operation, then $\Xi_\Phi$ is positive and
\begin{equation*}
d\, \ptr{\Xi_\Phi}=(\sum_{k} X_k^* X_k)^T\leq I \, .
\end{equation*}
The reverse of this statement is also
true, hence, any positive operator $\Xi$ on $\hi\otimes\hi$ induces an operation
provided that $d \, {\rm tr}_1[\Xi]\leq I$.
Thus, under the Choi-Jamiolkowski
isomorphism operations on $\hi$ are associated with a specific subset of
operators on $\hi\otimes\hi$. Unlike Kraus operators, the
Choi-Jamiolkowski operator $\Xi_\Phi$ for a given operation $\Phi$ is unique.
In terms of Choi-Jamiolkowski operators Eq.~\eqref{eq:operation->effect} reads
\begin{equation}
d\, \ptr{(\Phi\otimes\I)(\ket{\psi_+}\bra{\psi_+})}=A^T\, .
\end{equation}

For each operation $\Phi$, there is a minimal number of operators $X_k$ needed in its Kraus decomposition.
This number is called the \emph{(Kraus) rank} of $\Phi$, and we call any Kraus decomposition with this minimal number of elements a \emph{minimal Kraus decomposition}. (Notice, however, that even the choice of minimal Kraus decomposition is not unique.) Moreover, the rank of the associated
Choi-Jamiolkowski operator $\Xi_\Phi$ equals to the Kraus rank of the operation
$\Phi$, i.e. ${\rm rank}(\Phi)\equiv{\rm rank}(\Xi_\Phi)$.
Operations with Kraus rank 1 are called \emph{pure}. They are exactly the extremal elements in the convex set of all operations.

The following two classes of operations, namely, conditional state preparators and L\"uders operations, will be used later to exemplify the coexistence conditions.

\begin{example}\label{ex:preparator}
A \emph{conditional state preparator} is an operation $\Phi_A^{\xi}$
of the form
\begin{equation*}
\Phi_A^{\xi}(\varrho) = \tr{\varrho A} \xi
\end{equation*}
for some fixed $\xi\in\sh$ and $A\in\eh$. If $A=\id$,
then this operation is just the constant mapping $\varrho\mapsto\xi$.
For the conditional state preparator $\Phi_A^\xi$,
the associated Choi-Jamiolkowski operator $\Xi_A^\xi$ reads
\begin{eqnarray*}
\Xi_A^\xi&=&\frac{1}{d}\sum_{j,k}\Phi_A^\xi(\kb{\varphi_j}{\varphi_k})\otimes \kb{\varphi_j}{\varphi_k} \\
 &=& \frac{1}{d}\xi\otimes A^T\, .
\end{eqnarray*}
\end{example}

\begin{example}
Let $A$ be an effect. The \emph{L\"uders operation} $\Lu_A$ associated
to $A$ is defined by the formula
\begin{equation*}
\Lu_A(\varrho) = \sqrt{A} \varrho \sqrt{A} \, .
\end{equation*}
The associated Choi-Jamiolkowski operator $\Xi_A^\L$ is given by
\begin{equation}
\Xi_A^\L=\ket{(\sqrt{A}\otimes I)\psi_+}\bra{(\sqrt{A}\otimes I)\psi_+} \equiv
\ket{\psi_A}\bra{\psi_A}\,,
\end{equation}
with
\begin{equation*}
\tr{\Xi_A^\L}=\ip{\psi_A}{\psi_A}=\frac{1}{d}\tr{A}\leq 1\, .
\end{equation*}
If $A=\id$, then the corresponding L\"uders operation is the
identity operation $\I$ and we get
$\Xi_\id^\L=\ket{\psi_+}\bra{\psi_+}$.

Let us also notice that if $A$
is proportional to a one-dimensional projection $P$,
i.e., $A=\lambda P$ for some $0\leq\lambda\leq 1$,
then $\Lu_{A}$ is the conditional state preparator $\Phi_{A}^P$.
Namely, we have
\begin{equation*}
\sqrt{A}\varrho \sqrt{A}
=\lambda P\varrho P =\tr{\varrho \lambda P}P = \tr{\varrho A} P \, .
\end{equation*}
In all other cases, when rank $A>1$, L\"uders operation $\Lu_{A}$ is not a conditional
state preparator since for a conditional state preparator  $\Phi_{A}^\xi$ we have
${\rm rank}(\Xi^\xi_A)={\rm rank}(\xi){\rm rank}(A^T)\geq {\rm rank}(A)>1$, but
${\rm rank}(\Xi^\L_A)=1$.
\end{example}

\begin{figure}
\begin{center}
\includegraphics[scale=0.9]{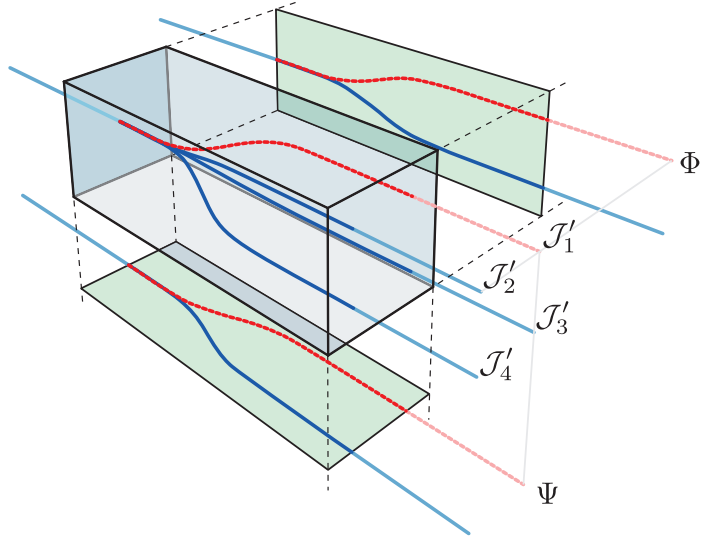}
\end{center}
\caption{\label{fig:coexistence_of_ops}Coexistence of two operations $\Psi$ and $\Phi$ is conditioned by the existence of a four-outcome instrument determined by four operations $\J'_1$ through $\J'_4$ such that $\J'_1+\J'_2=\Phi$ and $\J'_1+\J'_3=\Psi$.}
\end{figure}

We now make a simple but useful observation related to Def.~\ref{def:coexistent}.
Suppose that $\Phi$ and $\Psi$ are coexistent operations and that $\J$ is an instrument such that $\Phi,\Psi\in\ran\J$. This means that there are outcome sets $X,Y$ such that $\J_X=\Phi,\J_Y=\Psi$. We define another instrument $\J'$ with outcomes $\{1,2,3,4\}$ by setting
\begin{equation}
\J'_1 = \J_{X\cap Y} \, \quad \J'_2 = \J_{X\cap \neg Y} \, \quad \J'_3 = \J_{\neg X\cap Y} \, \quad \J'_4 = \J_{\neg X\cap \neg Y} \, .
\end{equation}
It follows from the properties of $\J$ that $\J'$ is indeed an instrument.
The operations $\Phi$ and $\Psi$ are in the range of $\J'$ as $\J'_1+\J'_2=\Phi$ and $\J'_1+\J'_3=\Psi$. Thus, we conclude the following.

\begin{proposition}\label{prop:basic}
Two operations are coexistent if and only if they are in the range of an instrument defined on the outcome set $\meo=\{1,2,3,4\}$.
\end{proposition}

The fact stated in Prop. \ref{prop:basic} simplifies the study of the coexistence relation as we need to concentrate only on four outcome instruments.
An illustration of a four outcome instrument and two coexistent operations is depicted in Fig.~\ref{fig:coexistence_of_ops}.

\begin{proposition}\label{prop:coex-kraus}
Two operations $\Phi$ and $\Psi$ are coexistent if and only if there exists a sequence of bounded operators $\{X_j\}_{j\in J}$ and index subsets $J_1,J_2\subseteq J$ such that
\begin{equation}\label{eq:general-coex-1}
\Phi(\cdot) = \sum_{j\in J_1} X_j \cdot X^\ast_j \, , \quad
\Psi(\cdot)= \sum_{j\in J_2} X_j \cdot X_j^\ast \, .
\end{equation}
and
\begin{equation}\label{eq:general-coex-2}
\sum_{j\in J} X_j^\ast X_j = \id \, .
\end{equation}
If $\Phi$ and $\Psi$ are coexistent, we can choose an index set $J$ with at most $3d^2+1$ elements.
\end{proposition}

\begin{proof}
Suppose first that there exists a sequence of bounded operators $\{X_j\}_{j\in J}$ with the required properties. By defining $\J_j (\rho)=X_j \rho X^\ast_j$ we get an instrument $\J$ having both $\Phi$ and $\Psi$ in its range. Hence, $\Phi$ and $\Psi$ are coexistent.

Suppose then that $\Phi$ and $\Psi$ are coexistent. As we have seen in Prop. \ref{prop:basic}, there exists a four outcome instrument $\J$ such that $\J_1+\J_2 =\Phi$ and $\J_1+\J_3=\Psi$. Choose a Kraus decomposition $\{X^{(k)}_j\}$ for each $\J_k$. The union $\cup_k \{X^{(k)}_j\}$ forms a collection with the required properties.

The last claim follows by noticing that each operation $\J_k, k=1,2,3,$ has a Kraus decomposition with (at most) $d^2$ Kraus operators. On the other hand, the role of the operation $\J_4$ is only to guarantee the normalization of the instrument $\J$. We can hence re-define $\J_4$ as the operation having a single Kraus operator $\sqrt{\id-\sum_{k=1}^3 \J_k^\ast(\id)}$.
\end{proof}

Let us note that the statement of Prop.~\ref{prop:coex-kraus} remains valid if the Eq.~\eqref{eq:general-coex-2} is replaced with an inequality
\begin{equation}\label{eq:general-coex-3}
\sum_{j\in J} X_j^\ast X_j \leq \id
\end{equation}
and then the number of the elements in $J$ can be chosen to be at most $3d^2$.
We can hence use either condition \eqref{eq:general-coex-2} or \eqref{eq:general-coex-3}, depending on which one happens to be more convenient.

In the following we formulate the basic coexistence criterion of Prop. \ref{prop:coex-kraus}  in terms of Choi-Jamiolkowski operators.

\begin{proposition}\label{prop:coex-cj}
Two operations $\Phi$ and $\Psi$ are coexistent if and only if
there exists a state $\Omega\in{\mathcal{S}}(\hi\otimes\hi)$ with
$d\, \ptr{\Omega}=\id$, which has a decomposition into four positive operators
$\Omega=\sum_{k=1}^4 \Xi_k$ such that
\begin{equation}
\Xi_\Phi=\Xi_1 + \Xi_2\, ,\quad
\Xi_\Psi=\Xi_1 + \Xi_3 \, .
\end{equation}
\end{proposition}

\begin{proof}
In the Choi-Jamiolkowski representation a four outcome instrument $\J$
translates into a mapping
$k\mapsto \Xi_k=(\J_k\otimes\I)(\kb{\psi_+}{\psi_+})$,
where $\Xi_k$ are positive operators on $\hi\otimes\hi$
and $\sum_{k=1}^4 d\, \ptr{\Xi_k}=\id$ meaning that
$\Omega \equiv\sum_{k=1}^4 \Xi_k$ is a state in ${\mathcal{S}}(\hi\otimes\hi)$.
The claim then follows from Prop. \ref{prop:basic}.
\end{proof}

\begin{example}\label{ex:unitary}
Let $U$ be a unitary operator and $\U$ the corresponding unitary channel, i.e., $\U(\varrho)=U\varrho U^\ast$. As $\U$ describes a deterministic and reversible state transformation, it is not expected to be coexistent with many other operations.
Of course, we can reduce $\U$ by accepting the state transformation with some probability $0\leq\lambda \leq 1$ and ignoring the rest, hence obtaining an operation $\lambda\U$. Thus, $\U$ and $\lambda\U$ are coexistent operations.

A proof that $\lambda\U$ are indeed the only operations coexistent with the unitary channel $\U$ can be seen from Prop. \ref{prop:coex-cj}.
The Choi-Jamiolkowski operator $\Xi_{\U}$ corresponding to $\U$ is $\kb{\psi_U}{\psi_U}$, where
\begin{equation*}
\psi_U=\frac{1}{\sqrt{d}}\sum_{j} U\varphi_j\otimes\varphi_j \, .
\end{equation*}
 In particular, $\Xi_{\U}$ is a one-dimensional projection and it can be written as a sum of two positive operators only if they are proportional to $\Xi_{\U}$.
On the other hand, there cannot be other operators in the decomposition of $\Omega$ as  $\Xi_{\U}$ is already normalized, $d{\rm tr}_1[\Xi_{\U}]=\id$ .
Therefore, an operation $\Phi$ is coexistent with $\U$ only if $\Xi_{\Phi}=\lambda\Xi_{\U}$ for some number $0\leq\lambda\leq 1$, which means that $\Phi=\lambda\U$.
\end{example}

\section{Trivial coexistence}\label{sec:rank1}

Let $\Phi$ and $\Psi$ be two coexistent operations. Referring to Prop.
\ref{prop:coex-kraus} we can meet with three possible situations:
\begin{itemize}
 \item[(C1)] $J_1\cap J_2 = \emptyset$;
 \item[(C2)] $J_1\subseteq J_2$ or $J_2\subseteq J_1$;
 \item[(C3)] none of the above.
\end{itemize}

In particular, if we can choose index subsets
$J_1,J_2$ such that $J_1\cap J_2=\emptyset$, then $\Psi+\Phi$ is also
an operation. Similarly, (C2) implies that $\Psi-\Phi$, or $\Phi-\Psi$
is an operation. It is clear that the verification of
the coexistence of $\Psi$ and $\Phi$ in such cases is straightforward.
We conclude that the coexistence of $\Phi$ and $\Psi$ falls out trivially if
\begin{itemize}
 \item[(T1)] $\Phi+\Psi$ is an operation;
 \item[(T2)] $\Phi-\Psi$ or $\Psi-\Phi$ is an operation;
\end{itemize}
thus, two operations satisfying one of the conditions (T1), (T2) are
called \emph{trivially coexistent}.

For a general coexistence problem we need to consider four outcome instruments;
however, in the case of trivial coexistence three outcome instruments are sufficient.
For instance, if $\Psi+\Phi$ is an operation, then we can choose a channel $\E$ such that also $\E-\Psi-\Phi$ is an operation and we can define a three outcome instrument $\J_1=\Phi$, $\J_2=\Psi$, $\J_3=\E-\Phi-\Psi$.
Similarly, if $\Psi-\Phi$ is an operation, we choose a channel $\E$ such that $\E-\Psi$ is an operation and we can define a three outcome instrument $\J_1=\Phi$, $\J_2=\Psi-\Phi$,  $\J_3=\E-\Psi$.

Let $\Phi_A$ and $\Psi_B$ be two operations. The condition (T1) means
\begin{equation*}
\tr{\Phi_A(\varrho)+\Psi_B(\varrho)}\leq 1
\end{equation*}
for every $\varrho\in\sh$.  This is equivalent to
\begin{equation*}
A+B\leq\id \, .
\end{equation*}

The set of operations is a partially ordered set if we adopt the following
relation between operations:
\begin{equation*}
\Psi\leq\Phi\quad\Leftrightarrow\quad \Phi-\Psi\ {\rm is\ operation}\,.
\end{equation*}
Hence, we can write (T2) as
\begin{equation*}
\Psi\leq\Phi\quad \textrm{or} \quad \Phi\leq\Psi \, .
\end{equation*}
Unlike (T1), these conditions do not reduce to generally valid effect inequalities in terms of the associated effects $A$ and $B$. Let us notice that a necessary condition for $\Phi_A\leq\Psi_B$ is that $A\leq B$. This is, however, not generally sufficient.  The form of (T2) depends on the operations in question, as we demonstrate in Examples \ref{ex:preparators-trivial} and \ref{ex:luders-trivial} below.

\begin{example}\label{ex:preparators-trivial}
Let $\xi$ be a fixed state, $A$ and $B$ two effects, and $\Phi_A^\xi$ and $\Phi_B^\xi$ the corresponding conditional state preparators. The trivial coexistence condition (T2) now becomes a requirement that either $A\geq B$ or $B\geq A$.
\end{example}

\begin{proposition}\label{prop:rank-1}
Let $\Phi$ and $\Psi$ be two pure operations. If $\Phi$ and $\Psi$ are coexistent, then they are trivially coexistent.
\end{proposition}

\begin{proof}
As $\Phi$ and $\Psi$ are pure, there are bounded operators $V,W$ such that $\Phi(\cdot)=V \cdot V^\ast$ and $\Psi(\cdot)=W \cdot W^\ast$. By Prop. \ref{prop:coex-kraus} the coexistence of $\Phi$ and $\Psi$ means that
\begin{equation*}
\Phi(\cdot) = \sum_{j\in J_1} X_j \cdot X^\ast_j = V \cdot V^\ast \, , \quad
\Psi(\cdot) = \sum_{j\in J_2} X_j \cdot X_j^\ast = W \cdot W^\ast \, .
\end{equation*}
It follows that $X_j = c_j V$ for every $j\in J_1$ and $X_j = d_j W$ for every $j\in J_2$. Here $c_j,d_j$ are non-zero complex numbers. This shows that we have two possibilities: either $J_1 \cap J_2 = \emptyset$ or $c_j V = d_j W$ for some $j \in J_1 \cap J_2$. The first case means that $\Phi+\Psi$ is an operation, while the second case leads to the condition $\Phi = p \Psi$ for some $p\in \real_+$. This implies that either $\Phi-\Psi$ or $\Psi-\Phi$ is an operation, depending whether $p\leq 1$ or $p\geq 1$.
\end{proof}

\begin{example}\label{ex:luders-trivial}
L\"uders operations are, by definition, pure operations and therefore their coexistence reduces to the trivial coexistence by Prop.~\ref{prop:rank-1}.
Different to Example \ref{ex:preparators-trivial}, the effect inequalities $A\leq B,B\leq A$ are
not sufficient to guarantee the coexistence of L\"uders operations $\Lu_A$ and $\Lu_B$.
For example, let $P$ be a one-dimensional projection. Then $P\leq I$, but
neither $\Lu_P-\Lu_I$, nor $\Lu_I-\Lu_P$, nor
$\Lu_I+\Lu_P$ are operations.
By Prop. \ref{prop:rank-1} we thus conclude that $\Lu_P$ and $\Lu_I$ are not coexistent
operations.

To see the content of the condition (T2), suppose that $\Lu_A-\Lu_B$ is an operation.
This implies that
\begin{equation*}
\kb{\sqrt{A}\psi}{\sqrt{A}\psi} \geq \kb{\sqrt{B}\psi}{\sqrt{B}\psi}
\end{equation*}
for every $\psi\in\hi$. As a consequence, $B=\lambda A$ for some $0\leq\lambda\leq 1$.

In summary, L\"uders operations are coexistent if and only if either $A+B\leq I$
or $A$ is proportional to $B$. The effects $A$ and $B$, for which the latter inequality holds, are called \emph{disjoint}.
\end{example}

\section{Coexistence of operations vs. coexistence of effects}\label{sec:effects}

As we have seen in Section \ref{sec:rank1} the trivial coexistence
of operations is closely related to the coexistence of effects.
In this section we investigate in more details the relations
between the coexistences of operations, their associated effects and
Choi-Jamiolkowski operators.

\begin{proposition}\label{prop:from_coex_operations}
If two operations $\Phi_A$ and $\Psi_B$ are coexistent, then the corresponding effects $A$ and $B$ are coexistent.
\end{proposition}

\begin{proof}
By Prop.~\ref{prop:coex-kraus} the coexistence of $\Phi_A$ and $\Psi_B$ is equivalent to the existence of a set $\{X_j\}_{j\in J}$ and
index subsets $J_1,J_2\subseteq J$ such that
\begin{equation*}
\Phi_A(\cdot)=\sum_{j\in J_1}X_j\cdot X_j^\ast,\qquad
\Psi_B(\cdot)=\sum_{j\in J_2}X_j\cdot X_j^\ast \, .
\end{equation*}
For each $j\in J$, we define $\G_j := X_j^\ast X_j$. The effects $\G_j$ define a discrete observable $\G$. The effects $A$ and $B$ belong to the range of $\G$ as $A=\sum_{j\in J_1}X_j^\ast X_j$ and $B=\sum_{j\in J_2}X_j^\ast X_j$. Therefore, $A$ and $B$ are coexistent.
\end{proof}

\begin{example}
Let $A$ and $B$ be two coexistent effects. This means that there is an observable $\G$ such that $\G(X)=A$ and $\G(Y)=B$. We fix a state $\xi$ and define an instrument $\J$ by formula
\begin{equation*}
\J_Z(\varrho) = \tr{\varrho \G(Z)} \xi \, .
\end{equation*}
The operations $\J_X$ and $\J_Y$ are then the conditional state preparators $\Phi_A^\xi$ and $\Phi_B^\xi$, respectively.
Thus, if $A$ and $B$
are coexistent, then also the conditional state preparations
$\Phi_A^\xi$ and $\Phi_B^\xi$ are coexistent.
\end{example}

As discussed in Section \ref{sec:general}, each operation $\Phi$ determines a Choi-Jamiolkowski operator
$\Xi_\Phi$ on $\hi\otimes\hi$ such that $d{\rm tr}_1[\Xi_\Phi]\leq I$.
As Choi-Jamiolkowski operators are effects on $\hi\otimes\hi$ we can formally consider their coexistence. Our aim is to investigate the relation between the coexistence of operations and the coexistence of
Choi-Jamiolkowski operators as effects.

If $\Phi$ and $\Psi$ are coexistent, then
the linearity of the Choi-Jamiolkowski isomorphism guarantees
that effects $\Xi_\Psi$ and $\Xi_\Phi$ are coexistent, too. However,
the converse is not true. Namely, even if $\Xi_\Phi$ and $\Xi_\Psi$
are coexistent as effects, the associated operations $\Phi,\Psi$ need not be
coexistent. For example, according to Proposition \ref{prop:rank-1} two rank-1
operations $\Phi_A$ and $\Psi_B$ are coexistent only if they are trivially
coexistent. However, if $A,B$ are one-dimensional projections associated
with vectors $\varphi,\eta\in\hi$, then
$\Xi_A=\frac{1}{d}\ket{\varphi}\bra{\varphi}
\otimes(\ket{\varphi}\bra{\varphi})^T$ and
$\Xi_B=\frac{1}{d}\ket{\eta}\bra{\eta}
\otimes(\ket{\eta}\bra{\eta})^T$, which are always (trivially)
coexistent as effects,
because $\Xi_A+\Xi_B\leq I\otimes I$. The point is that $I\otimes I$
does not correspond to any operation, because
$d{\rm tr}_1[I\otimes I]=d^2I\not\leq I$.

Table~\ref{tab:relations} summarizes the mentioned results.
We see that the remaining problem is the following: if $A$ and $B$ are coexistent effects but do not satisfy $A+B\leq\id$, what are the coexistent operations $\Phi_A$ and $\Psi_B$? The following examples demonstrate different aspects of this general problem.

\renewcommand\arraystretch{2}
\begin{table}
\begin{tabular}{l@{$\quad$}c@{$\quad$}p{5cm}}
\hline
$A$ and $B$ are not coexistent & $\Longrightarrow$ & there are no coexistent operations $\Phi_A$ and $\Psi_B$
\\[1ex]
\hline
$A$ and $B$ are coexistent & $\Longrightarrow$ & there exist coexistent operations $\Phi_A$ and $\Psi_B$
\\[1ex]
\hline
$A+B\leq\id$  &  $\Longrightarrow$ & all operations $\Phi_A$ and $\Psi_B$ are trivially coexistent
\\[1ex]
\hline
$\Xi_\Phi$ and $\Xi_\Psi$ are not coexistent &  $\Longrightarrow$ &  $\Phi$ and $\Psi$ are not coexistent\\[1ex]
\hline
\\
\end{tabular}
\caption{\label{tab:relations} Relations between coexistence of effects and coexistence of their compatible operations.}
\end{table}
\renewcommand\arraystretch{1}

\begin{example}
Let $\Phi_A^{\xi_1}$ and $\Phi_B^{\xi_2}$ be two conditional state preparators such that $\xi_1$ and $\xi_2$ are pure states, i.e., $\xi_i=\kb{\phi_i}{\phi_i}$ for some unit vectors $\phi_i\in\hi$.
A Kraus decomposition of $\Phi_A^{\xi_1}$, with Kraus operators $X_k$, is of the form $\kb{\phi_1}{\eta_k}$ for some vector $\eta_k$, or a sum of these kind of operators.
Similarly, a Kraus decomposition of $\Phi_B^{\xi_2}$, with Kraus operators $Y_l$, is either $\kb{\phi_2}{\eta'_l}$ for some vector $\eta'_l$, or a sum of these kind of operators.
Suppose  that $\Phi_A^{\xi_1}$ and $\Phi_B^{\xi_2}$ are coexistent, but the inequality $A+B\leq\id$ does not hold.
This implies that $X_k=Y_l$ for some indices $k$ and $l$. As a consequence, we must have $\xi_1=\xi_2$.
We conclude that if $\xi_1\neq\xi_2$ then the conditional pure state preparators are coexistent only if $A+B\leq I$.
\end{example}

\begin{example}
It is customary to call an effect $A$ \emph{trivial} if it is of the form $\lambda\id$ for some $0\leq \lambda \leq 1$.
Trivial effects are exactly those effects which are coexistent with all the other effects.

In the same way, we can call an operation trivial if it is coexistent with all the other operations.
Clearly, the null operation $\Lu_{\nul}(\varrho)=\nul$ is trivial in this sense since any instrument can be expanded by adding one additional outcome and attaching $\Lu_{\nul}$  to this additional outcome.

Actually, the null operation is the only trivial operation. As shown in Example \ref{ex:unitary} a unitary channel $\U$ is coexistent only with operations $\lambda\U$.
Since a trivial operation is coexistent with all unitary channels, it must be the null operation.
\end{example}

\section{Discussion}\label{sec:discussion}

In this paper we have studied the coexistence of two quantum operations.
In particular, we have shown that two common types of operations in quantum information, namely conditional state preparations and L\"uders operations, are coexistent only under some very restrictive conditions. We have also shown that the coexistence problem for operations does not reduce to the coexistence problem for effects.

Recently, coexistence of two arbitrary qubit effects has been characterized \cite{StReHe08,BuSc09,YuLiLiOh08,LiLiYuCh09}. It would be interesting to give an analogous characterization of two arbitrary qubit operations. This problem, however, seems to be much more intricate as already the parametrization of the qubit operations is quite a complex task \cite{RuSzWe02}.

In quantum information theory, it has become typical to consider impossible devices, forbidden by the rules of quantum mechanics. For an impossible device, one can then study its best approximative substitute. Especially, we can ask for the best coexistent approximations for two non-coexistent L\"uders operations. This problem will be studied elsewhere.

\section*{Acknowledgements}
T.H. acknowledge financial support from QUANTOP and Academy of Finland.
D.R., P.S. and M.Z. acknowledge financial support via the European Union project
HIP FP7-ICT-2007-C-221889, and via the projects
APVV-0673-07 QIAM, OP CE QUTE ITMS NFP 262401022,
and CE-SAS  QUTE. M.Z. acknowledges also support of GA\v CR via
project GA201/07/0603.


\end{document}